\documentclass[twocolumn,english,prl]{revtex4}
\usepackage[latin9]{inputenc}
\usepackage{amsmath}
\usepackage{amssymb}
\usepackage{graphicx}
\usepackage{esint}

\makeatletter
\@ifundefined{textcolor}{}
{%
 \definecolor{BLACK}{gray}{0}
 \definecolor{WHITE}{gray}{1}
 \definecolor{RED}{rgb}{1,0,0}
 \definecolor{GREEN}{rgb}{0,1,0}
 \definecolor{BLUE}{rgb}{0,0,1}
 \definecolor{CYAN}{cmyk}{1,0,0,0}
 \definecolor{MAGENTA}{cmyk}{0,1,0,0}
 \definecolor{YELLOW}{cmyk}{0,0,1,0}
}

\usepackage{babel}

\makeatother

\usepackage{babel}
\begin{document}

\title{Magnetic field-dependent inhomogeneities and their effect on the
magnetoresponse of 2D superconductors}

\author{S. Sankar}

\author{V. Tripathi}

\affiliation{Department of Theoretical Physics, Tata Institute of Fundamental
Research, Homi Bhabha Road, Navy Nagar, Mumbai 400005, India}

\date{\today}
\begin{abstract}
We show that inhomogeneities in the spatial distribution of Cooper
pairs and in the phase of the local superconducting order parameter
in the vicinity of a superconductor-normal state transition (SNT)
in two dimensions can be highly sensitive to a perpendicular magnetic
field. We focus on the role of orbital effects in the field-dependence
of local superfluid stiffness and superconducting phase disorder in
homogeneously-disordered two-dimensional superconductor thin films.
The relative importance of these orbital effects is analyzed in different
physical regimes dominated by Coulomb blockade, thermal phase fluctuations
and Aharanov-Bohm phase disorder respectively. Following this approach,
we obtain explicit expressions for the field dependence of magnetoresistance
and superfluid stiffness near the SNT, and attempt an understanding
of some recent experimental findings. 
\end{abstract}
\maketitle
One of the most challenging problems in strongly disordered superconductors
relates to understanding the nature of the magnetic field-induced
superconductor-normal state transition (SNT). Experimental and theoretical
studies over the past two decades have opened a large number of puzzling
questions such as the origin of the giant non-monotonous magnetic field
dependence of the resistivity \cite{sambandamurthy,kapitulnik,dolgopolov,yigal,pokrovskyprl,senthil,vinokur_hopping},
flux quantization in the insulating state \cite{stewart} and the
universality class governing the field-induced SNT \cite{kapitulnik2016,paalanen1992,fisher1990,dolgopolov}.
The two-dimensional (2D) case has in particular attracted intense
theoretical attention and it is the focus of this work. In the absence
of a magnetic field, it is well-known that strong homogeneous disorder
introduces granularity in the form of superconducting islands embedded
in an insulating matrix \cite{ioffe1981,fisher1989,ghosal1998,sacepe2008,chand2012}.
However the role of an external magnetic field on the SNT through
its effect on the distribution of such islands \cite{pokrovskyprl,yigal}
and on the associated phase frustration brought in by the Aharonov-Bohm
(AB) phases of the Cooper pairs tunneling across the islands \cite{dasgupta1988,kim2008quantum}
is not well-understood and is a topic of considerable current debate.

Mean-field analyses of the field-sensitivity of the distribution of
superconducting regions go back nearly two decades for weakly-disordered
metals \cite{spivak1995,galitski2001}, and more recently, \cite{yigal}
for strongly-disordered insulators. Standard, perturbative approaches
fail in the strongly-disordered regime but numerical mean-field solutions
of the appropriate Bogoliubov -- de Gennes (BdG) equations \cite{yigal}
reveal a picture of shrinking superconducting regions in increasing
fields, a downward shift of the distribution of the local superconducting
gaps, and through the Ambegaokar-Baratoff relation \cite{ambegaokarbaratoff},
a corresponding decrease in the Josephson couplings $J$ between neighboring
grains. To understand the physical origin of these effects, we study
a phenomenological model of repulsive bosons (Cooper pairs) subjected
to a disordered potential and a perpendicular magnetic field. The
approach is reminiscent of earlier work on Lifshitz states \cite{lifshitz}
in disordered Bose systems \cite{larkin1995,falcoprb,pokrovskyprl}.
We obtain the typical size and separation of the superconducting islands
and show that wave function shrinking in the presence of a magnetic
field suppresses the Josephson couplings as $J(B)\sim\exp[-(B/B_{J})^{2}].$ 

To understand magnetoresponse of these 2D granular superconductors,
we study the standard Josephson-junction (XY) model, 
\begin{align}
L & =\frac{1}{4E_{c}}\sum_{\mathbf{i}}(\partial_{\tau}\phi_{\mathbf{i}})^{2}-\sum_{\langle\mathbf{ij}\rangle}J_{\mathbf{ij}}(B)\cos(\phi_{\mathbf{ij}}+A_{\mathbf{ij}}),\label{eq:model}
\end{align}
where $E_{c}$ represents the Coulomb blockade scale, $\phi_{\mathbf{ij}}=\phi_{\mathbf{i}}-\phi_{\mathbf{j}}$
is the superconducting phase difference between neighboring grains
at positions $\mathbf{i}$ and $\mathbf{j}$ respectively, and $A_{\mathbf{ij}}=(2e/\hbar)\int_{\mathbf{i}}^{\mathbf{j}}\mathbf{A}\cdot d\mathbf{r}$
are the AB phases acquired by the hopping Cooper pairs. Disregarding
the contribution of normal quasi particles means the model can provide
a good description of the magnetoresponse only at lower fields where
Cooper pair breaking is not important. Spatial disorder in the grain
positions introduces randomness in the Josephson couplings as well
as the AB phases. Studies of the 2D classical limit of Eq.(\ref{eq:model})
in the $B=0$ limit \cite{yigal} have shown that strong disorder
in $J$ does not alter the universality class of the SNT from the
homogeneous case (where it is known to be of Kosterlitz-Thouless (KT)
type) but is nevertheless dominated by a percolating backbone of paths
with the largest local superfluid stiffnesses. Likewise the transition
in the quantum 1D disordered counterpart at $B=0$ also falls in the
KT universality class \cite{altman,giamarchi}. Therefore for simplicity
we will work with the typical value of $J$ ignoring its spatial disorder.

In regular lattices, the AB phase is associated with flux threading
the plaquettes, and depending on the amount of frustration $f$ (measured
as a fraction of a flux quantum), leads to oscillations in properties
such as the critical current and the resistance \cite{teitel1983josephson,van1996quantum}.
Such matching (commensuration) effects are absent in the disordered
case as there is random flux penetration in different plaquettes.
In a phenomenal work, Carpentier and Le Doussal \cite{carpentiernucphysb,carpentierprl}
studied phase transitions in the classical quenched random phase XY
model on a square lattice close to integer $f.$ The presence of disorder
results in rare favorable regions for the occurrence of vortices at
low temperatures. At sufficiently low temperatures, they found that
the disorder-induced phase transition is not in the KT universality
class. Very similar results were also obtained earlier \cite{giamarchi}
in a study of the Anderson localization in one-dimensional Luttinger-liquids
subjected to quenched phase disorder. The similarity is puzzling since
quenched disorder in 1D is equivalent to columnar disorder in the
two-dimensional case. Quantum Monte Carlo studies \cite{kim2008quantum}
of the interplay of phase frustration and Coulomb blockade suggest
a zero temperature field-driven SNT with dynamic exponent $z\approx1.3,$
placing the transition in a different universality class from 3D XY. 

In this Letter we study the effect of three dominant mechanisms governing
loss of phase coherence and their specific signatures on the magnetoresistance
and superfluid stiffness. These are (a) quantum phase fluctuations
originating from Coulomb blockade, (b) thermal fluctuations of the
phase and (c) frustration effects due to disorder in AB phases. We
show that Coulomb blockade effects impart a specific signature to
the magnetoresistance, $\rho(B)\sim\exp[(B/B_{0})^{2}].$ Where the
SNT is driven by thermal fluctuations, we find a KT transition, with
$\rho(B)\sim\exp[-1/\sqrt{B-B_{KT}}]$ in the critical region. In
the AB phase frustration dominated regime, we find a new, non KT critical
behavior, $\rho(B)\sim\exp[-1/(B-B_{AB})].$ The field-dependent superfluid
stiffness $\Upsilon$ also shows a surprising behavior: at small fields,
we find that phase frustration effects on $\Upsilon$ are more significant
than the field dependence of Josephson couplings. In the Coulomb blockade
regime away from the critical region, our predicted magnetoresistance
is in excellent accord with experimental data \cite{sambandamurthy,kapitulnik}.
However in the critical scaling region, existing experimental data
is somewhat less clear, and while there is some evidence for mechanism
(c) for the field-tuned SNT in oxide heterostructures\cite{budhani}
, further study is needed and we propose additional probes to distinguish
between the two.

\begin{figure}
\includegraphics[width=7cm]{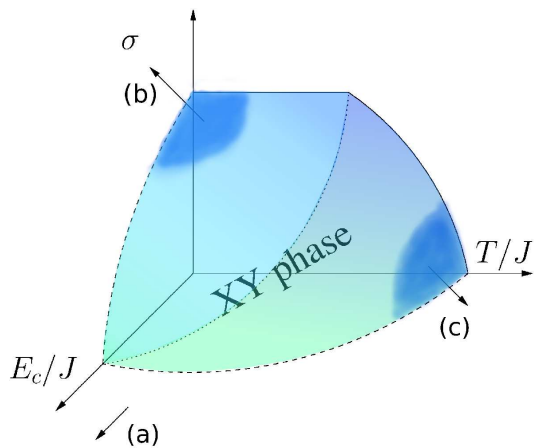}

\caption{\label{fig:phase-diag}Schematic phase diagram for the 2D superconductor-normal
state transition, with the XY superfluid phase in the interior of
the surface, as a function of dimensionless temperature $T/J,$ Coulomb
blockade scale $E_{c}/J$ and the Aharanov-Bohm (AB) phase disorder
$\sigma$ for the model described in Eq. (\ref{eq:model}). Here $J$
is the (field-dependent) Josephson coupling estimated in the paper.
(a) refers to the Coulomb blockade dominated regime. The shaded regions
(b) and (c) denote transitions driven by AB phase frustration and
thermal phase fluctuation respectively. The dotted line on the surface
separates these two different critical scaling regimes. The critical
disorder at low $T$ and $E_{c}$ is independent of $T$ and scaling
of the correlation length is not of the Kosterlitz-Thouless type \cite{carpentiernucphysb}. }
\end{figure}

We now analyze the effect of a transverse magnetic field on the distribution
of the SC islands in the granular superconductor. Consider a model
of repulsive bosons (Cooper pairs) with average density $n$ subjected
to a random potential with a Gaussian white noise distribution:
\begin{align}
H & =\sum_{\mathbf{p}}\frac{\Pi^{2}}{2m}a_{\mathbf{p}}^{\dagger}a_{\mathbf{p}}+\int_{\mathbf{r}}\left[\frac{g}{2}|\Psi(\mathbf{r})|^{4}+U(\mathbf{r})|\Psi(\mathbf{r})|^{2}\right],\label{H_Bose}
\end{align}
where $\Psi(\mathbf{r})=\frac{1}{\sqrt{V}}\sum_{\mathbf{p}}a_{\mathbf{p}}\exp[i\mathbf{p}\cdot\mathbf{r}/\hbar],$
$\Pi=(\mathbf{p}-q\mathbf{A}),$ $U(\mathbf{r})$ is the random potential,
$\langle U(\mathbf{r})\rangle=0$ and $\langle U(\mathbf{r})U(\mathbf{r}')\rangle=\kappa^{2}\delta(\mathbf{r}-\mathbf{r}'),$
$q=2e$ is the boson charge and $g,$ parametrizes the boson repulsion.
We choose the gauge $\mathbf{A}=\frac{1}{2}(\mathbf{B}\times\mathbf{r})$
with the field in the transverse $z$ direction. This model is equivalent
to earlier studied (for $B=0$) Ginzburg-Landau models with disorder
in critical temperature \cite{ioffe1981}. The important length scales
in the model are the single particle localization length ${\cal L}=\hbar^{2}/m\kappa$
characterizing the disorder, and the magnetic length $l_{B}\sim\sqrt{e/(2\pi\hbar B}).$
We are specifically interested in the regime ${\cal L}/l_{B}\ll1.$
At low densities, the interplay of disorder and interpaticle repulsion
leads to the formation of disconnected islands of localized bosons
\cite{falcoprb} whose typical size and separation may be estimated
as follows. The optimal potential fluctuation that has a bound state
at energy $E<0$ is found by minimizing $\frac{1}{2}\int U^{2}d\mathbf{r}+\lambda(E-H),$
where $\lambda$ is a Lagrange multiplier. We choose $\Psi$ to be
real, assuming a spherical fluctuation and zero angular momentum bound
state. Varying with respect to $U,$ we obtain $U=\lambda\Psi^{2};$
thus the size $R$ of the optimum potential well is also of the same
order as the wave function. The energy of a particle in an island,
in the mean-field approximation, is thus of the order of $-\frac{\hbar^{2}}{2mR^{2}}+\frac{(qBR)^{2}}{8m}+gN_{p}/(\pi R^{2}),$
where $N_{p}$ is the number of bosons in the island. The density
$n_{w}=n/N_{p}$ of these islands is determined by the Gaussian factor,
$\exp[-\frac{1}{2\kappa^{2}}\int_{\mathbf{r}}U^{2}],$ whence $n_{w}\sim(1/\pi R)^{2}\exp[-(\hbar^{2}/m\kappa R)^{2}]=(1/\pi R)^{2}\exp[-({\cal L}/R)^{2}].$
Minimizing the energy with respect to $R,$ the size of the typical
island, to logarithmic precision, is
\begin{align}
R(B) & \sim\frac{{\cal L}}{\sqrt{\ln[n_{c}(B)/n]}},\label{eq:RB}
\end{align}
where $n_{c}(B)\approx\frac{\hbar^{2}}{2gm{\cal L}^{2}}\left(1+\frac{(qB{\cal L}^{2}/\hbar)^{2}}{4}\frac{1}{\ln^{2}(n_{c}(0)/n)}\right)$
for small fields, is the critical density for percolation of the islands
and $n_{c}(B)/n>1.$ For future convenience we introduce $w(B)=n_{c}(B)/n.$
Clearly the magnetic field shrinks the islands but the field-dependence
is very different from a simple expectation from wave function shrinking
of a localized noninteracting particle. The distance $D\sim1/\sqrt{n_{w}}$
between the islands can be estimated as $D(B)\sim R(B)e^{\frac{1}{2}({\cal L}/R(B))^{2}}\sim{\cal L}\sqrt{w(B)},$
whence, on account of the exponential dependence of inter-island tunneling
probability, the typical inter-island Josephson coupling behaves as
$J\sim e^{-2D(B)/R(B)}$ 
\begin{align}
J(B) & \sim e^{-2\sqrt{w(B)}}.\label{eq:JB}
\end{align}
Note that even when at small magnetic fields, \textbf{${\cal L}/l_{B}\ll1,$}
the exponent in Eq. \ref{eq:JB} can be large at low boson densities,
$w(B)\gg1.$ For such fields we have $J(B)/J(0)\sim e^{-(B/B_{J})^{2}},$
where $B_{J}^{-2}\approx\frac{\sqrt{w(0)}}{\ln^{2}w(0)}(q{\cal L}^{2}/2\hbar)^{2}.$
We now analyze the effects of the three different mechanisms that
lead to loss of global phase coherence in their regimes of dominance
which are determined by the dimensionless parameters $E_{c}/J,$ $T/J$
and $\sigma,$ with the latter a measure of disorder in the fluxes
through elementary plaquettes. Figure \ref{fig:phase-diag} shows
the phase diagram and the regimes of our study. 

\emph{(a) Quantum phase fluctuations dominated insulating regime ($E_{c}/J,\, E_{c}/T\gg1$)}
: We treat the Josephson term in Eq.(\ref{eq:model}) as a perturbation,
and calculate the conductivity using the Kubo formula \cite{vikram_thermal,efetov2009}.
Transport in this model proceeds through Arrhenius activation and
incoherent sequential hopping of charges between neighboring islands
- this leads to a resistivity of the form 
\begin{align}
\rho(B) & \sim J(B)^{-2}e^{E_{c}(B)/T}\sim e^{[4\sqrt{w(B)}+(q^{2}/{\cal L}T)\sqrt{\ln w(B)}]}.\label{eq:rho-CB}
\end{align}
The above behavior shows the insulating nature of the normal state.
For small fields, the magnetoresistance obeys the law $\rho(B)/\rho(0)\sim\exp[(B/B_{0})^{2}],$
where\textbf{ $B_{0}^{-2}\approx\frac{(q{\cal L}^{2}/2\hbar)^{2}}{2\ln^{2}w(0)}[\sqrt{w(0)}+\frac{q^{2}}{{\cal L}T\sqrt{\ln w(0)}}].$}
More accurately, one must also take into account the renormalization
of the charging energy by Josephson coupling \cite{efetov2009,fistul2008},
$E_{c}\rightarrow E_{c}-J.$ It is interesting to note that a similar
field dependence of resistivity $\rho(B)\sim e^{(B/B_{c}-1)^{2}}$
has been obtained in the context of a superconductor to Hall insulator
transition \cite{shimshoni1998}.

\emph{(b) AB phase frustration dominated regime ($E_{c}/J\ll1,\, T/J\ll1,\,\sigma/\sigma_{c}\sim1$)}:
To study this regime, it is useful to consider the Coulomb gas representation
of the model in Eq. (\ref{eq:model}). Following earlier works \cite{carpentiernucphysb,petkovic2009}
we assume a Gaussian white noise distribution for the AB phases on
the links, reckoned from a background average corresponding to a typical
separation of islands, $D.$ In the Coulomb gas representation, such
disorder translates to a random flux threading elementary plaquettes,
corresponding to an external potential $V_{\mathbf{r}}$ acting on
the ``charges'' (vortices) with a Gaussian distribution $\langle(V_{\mathbf{r}}-V_{\mathbf{r}'})^{2}\rangle=4\sigma J^{2}\ln|\mathbf{r}-\mathbf{r}'|+O(1).$
Denoting the plaquette area fluctuation by $(\delta D)^{2},$ we identify
$\sigma\sim B^{2}(\delta D)^{4}.$ It is crucial that the random background
potential has long-range (logarithmic) correlations. In the continuum
description of the model with a lower cutoff scale $a_{0},$ $V_{\mathbf{r}}$
has a local part $v_{\mathbf{r}}:\,\langle(v_{\mathbf{r}}-v_{\mathbf{r}'})^{2}\rangle\sim\sigma J^{2}$
and a long-range correlated part $V_{\mathbf{r}}^{>}$ with no cross-correlation
between these two parts. The Coulomb gas Hamiltonian then reads 
\begin{align}
H & =-J\sum_{\mathbf{r}\neq\mathbf{r}'}n_{\mathbf{r}}n_{\mathbf{r}'}\ln\left(\frac{|\mathbf{r}-\mathbf{r}'|}{a_{0}}\right)-\sum_{\mathbf{r}}\left[n_{\mathbf{r}}V_{\mathbf{r}}^{>}-\ln Y[n_{\mathbf{r}},\mathbf{r}]\right],\label{eq:H-CB}
\end{align}
where $n_{\mathbf{r}}$ represents integer charge at $\mathbf{r}$
and the spatially dependent fugacities have the bare value, $\ln Y[n_{\mathbf{r}},\mathbf{r}]=\gamma Jn_{\mathbf{r}}^{2}+n_{\mathbf{r}}v_{\mathbf{r}},$
and $\gamma$ is a constant of order unity. We have dropped the background
term as it just sets the chemical potential of the vortices and does
not affect the scaling equations \cite{sondhi}.

In the absence of disorder, the usual RG procedure consists of (i)
increasing the short scale cutoff, $a_{0}\rightarrow a_{0}+dl,$ and
eliminating all dipoles in the annulus of thickness $dl,$ and (ii)
disregard all configurations that increase the net charge within the
cutoff region. The RG procedure is perturbatively controlled by small
dipole fugacities. For the disordered case, we follow Ref.\cite{carpentiernucphysb}
and introduce replicas which allows us to perform the average over
Gaussian disorder. The lowest excitations continue to carry charges
$0,\,\pm1$ but now the $n_{\mathbf{r}}^{\alpha}$ also carry a replica
index $\alpha.$ An important difference from the RG procedure of
the disorder-free case is that now when the cutoff is increased, one
must, apart from considering annihilation of replica charges, also
take into account ``fusion'' of unit charges in different replicas
(see appendix). Another important difference that invalidates the usual
perturbative expansion in small dipole fugacities is that the random
potential creates favorable regions for single vortex formation. Hence
we study the scale dependence of the single vortex fugacity distribution
identifying the density of rare favorable regions, $\rho_{a_{0}}^{v},$
for the occurrence of vortices as the perturbation parameter. By studying
the scaling of $\rho_{a_{0}}^{v},$ two distinct regimes can be identified
for $T/J\ll1$: (a) an XY phase phase at sufficiently low bare disorder
where $\rho_{a_{0}}^{v}$ scales to zero, and (b) a disordered phase
beyond a critical bare disorder where $\rho_{a_{0}}^{v}$ diverges
(see appendix for details). In the disordered phase, the phase correlation
length has a surprising non-KT behavior, $\xi\sim e^{1/(\sigma-\sigma_{c})},$
which in our context translates to a field dependence $\xi\sim e^{1/(B-B_{AB})},$
with $B_{AB}\sim\hbar/q(\delta D)^{2}.$ Such a non-KT behavior is
a direct consequence of the logarithmic scaling of the disorder potential
correlations. Another peculiarity is that over a range of low temperatures
up to a scale of order $J,$ the critical disorder $\sigma_{c}$ is
independent of the temperature \cite{carpentiernucphysb}. 

We obtain the magnetic field dependence of the superfluid stiffness
by solving the scaling equations in the critical region at low temperatures
for the coupling constant $J_{l}$ and the effective disorder $\sigma_{l}.$
Taking the ratio of the scaling equations for $J_{l}$ and $\sigma_{l}$
obtained in Ref.\cite{carpentiernucphysb}, we get 
\begin{align*}
\frac{\partial_{l}J_{l}^{-1}}{\partial_{l}\sigma_{l}} & \sim\frac{1}{J_{l}\sqrt{\sigma_{l}}},
\end{align*}
 and from the solution $J_{l}\sim e^{-2\sqrt{\sigma_{l}}}$ it follows
that the superfluid stiffness $\Upsilon(B)$ has the behavior 
\begin{align}
\Upsilon(B) & \sim J(B)e^{-2\sqrt{\sigma(B)}}\sim e^{-(B/B_{1})-(B/B_{J})^{2}},\label{eq:AB-stiffness}
\end{align}
where $B_{1}$ is of the order of $B_{AB}.$ Phase frustration effects
thus play a more important role in determining the low-field dependence
of superfluid stiffness in the AB phase-frustration dominated regime
compared to the effect coming from orbital shrinking.

Now we analyze magnetoresistance in the disordered phase at low temperatures
and close to the field-induced transition. Following Halperin and
Nelson \cite{halperin1979} we estimate the electrical resistivity
(which is essentially the vortex conductivity) as $\rho(B)=\mu_{v}n(B),$
where $\mu_{v}$ is the temperature and field-dependent mobility of
the vortices, and $n(B)\sim1/\xi^{2}$ is the vortex density. We make
an assumption that $\mu_{v}(B)$ is well-behaved near $B=B_{AB},$
which allows us to neglect its field dependence in comparison to the
singular behavior of $\xi(B).$ The temperature dependence of resistivity
is governed by the temperature dependence of the mobility, and we
believe it shows an activated behavior given the logarithmic Coulomb
interaction of the vortices\cite{shklovskii2008}. The magnetoresistance
in this AB phase frustration dominated regime thus grows as 
\begin{align}
\rho(B) & \sim\mu_{v}(T)e^{-1/(B-B_{AB})}.\label{eq:AB-rho}
\end{align}

\emph{(c) Thermal phase fluctuations dominated KT regime ($E_{c}/J\ll1,\,\sigma/\sigma_{c}\ll1,\, T/J(B)\sim1$)}:
In this regime, the transition is brought about by the proliferation
of thermally activated vortices. The superfluid stiffness now has
a field dependence$\Upsilon(B)\sim J(B)\sim e^{-(B/B_{J})^{2}}$ arising
from orbital shrinking of the superconducting islands. For the resistivity
we again consider the correlation length in the disordered phase,
which has the well-known form, $\xi\sim e^{1/\sqrt{T-T_{KT}}},$ with
$T_{KT}\propto J(B).$ Near the transition, this is equivalent to
a field-dependent correlation length, $\xi\sim e^{1/\sqrt{B-B_{KT}}}.$
Thus the resistivity in this regime has the form 
\begin{align}
\rho(B) & \sim\mu_{v}(T)e^{-1/\sqrt{B-B_{KT}}}.\label{eq:KT-rho}
\end{align}
For regimes (b) and (c), the normal state has a ``metallic'' temperature
dependence since enhancement of vortex mobilities at higher temperatures
translates to higher resistivity.

\begin{figure}
\includegraphics[width=7cm]{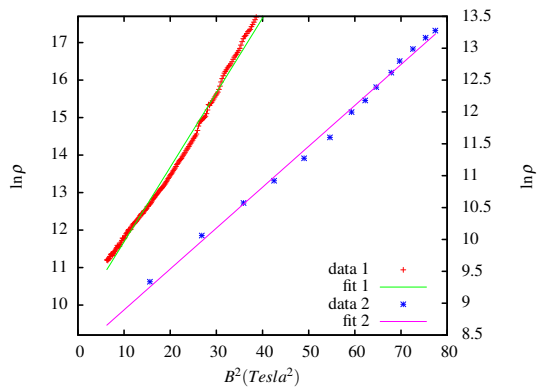}

\caption{\label{fig:data}Resistivity $\rho$ as a function of perpendicular
magnetic field $B$ for disordered InO$_{x}$ thin films as reported
in Ref.\cite{sambandamurthy} (data 1) and Ref. \cite{kapitulnik}
(data 2) in the low-field region where the fits are to the predicted
behavior $\rho(B)\sim e^{(B/B_{0})^{2}}$ corresponding to the Coulomb
blockade dominated regime discussed in the text.}
\end{figure}

\emph{Relation to experiments}: Figure \ref{fig:data} shows the low-temperature
and low-field magnetoresistance of disordered InO$_{x}$ thin films
extracted from two different experiments \cite{sambandamurthy,kapitulnik}.
The positive magnetoresistance data is very well-described by Eq.
(\ref{eq:rho-CB}) which places these samples in our Coulomb blockade
dominated regime. Deviation from the Coulomb blockade prediction is
seen near the magnetoresistance peak and we believe this is due to
the quasi particle transport channel opening up. In samples with lower
disorder \cite{kapitulnik}, unsurprisingly, Coulomb blockade does
not adequately explain the data; however, the other critical scaling
regimes (AB phase frustration and KT) show better agreement even though
we were unable to distinguish between the two (see appendix). In a recent
study of the field-tuned SNT at 2D interfaces of gated oxide heterostructures
\cite{budhani}, it was reported that for certain gate voltages, the
critical magnetic field at low temperatures was independent of the
temperature, suggestive of the phase frustration driven SNT mechanism.
Finally, our predictions for superfluid stiffness in the XY regime
can possibly be tested through studies of field-dependent ac conductivity\cite{misra2013}
and may provide an independent means for distinguishing between the
two regimes in the XY phase. 

In summary, we studied the field-dependence of the distribution of
SC islands in strongly disordered superconductors and constructed
an effective Josephson-junction model with field-dependent parameters.
Analyzing the model in different physical regimes - dominated by Coulomb
blockade, thermal phase fluctuations or Aharanov-Bohm phase fluctuations
- we obtained the field-dependence of resistivity and superfluid stiffness.
In the Coulomb blockade regime, available experimental data is in
excellent agreement with our prediction $\rho(B)\sim e^{(B/B_{0})^{2}},$
while in the critical scaling region, available magnetoresistance
data \cite{kapitulnik} is insufficient to distinguish between KT
and AB phase frustration regimes. 

At very low temperatures, the critical behavior in the vicinity of
the quantum critical point ($E_{c}/J(B)\sim1$) is expected to be
that of the 3D XY universality class. For the field-tuned transition
in systems with homogeneous potential disorder, the rapid decrease
of the Josephson coupling $J(B)$ with field implies that the likely
experimental trajectories in the $T/J\,\mbox{vs.}E_{c}/J$ plane rapidly
move out of the quantum critical region into the Coulomb-blockade
dominated region where $E_{c}/J\gg1.$ In contrast, in systems such
as nanopatterned superconducting proximity arrays, the fabrication
technique is such that the separation of superconducting regions (and
thus $J(B)$) is not as field-sensitive. Such systems look attractive
from the point of view of studying the critical behavior near the
field-tuned SNT, especially in the quantum critical region $E_{c}/J(B)\sim1.$
In our study we neglected pair-breaking effects which likely play
a crucial role in explaining the giant negative magnetoresistance
observed at higher fields \cite{pokrovskyprl,yigal}. Pair breaking
opens up an additional quasi particle transport channel, and it would
be interesting to study magnetic field effects in phase models with
both quasi particle and Cooper pair tunneling. 
\begin{acknowledgments}
We are grateful to G. Sambandamurthy for sharing his experimental
data with us, and to Y. Meir, P. Raychaudhuri, E. Shimshoni for valuable
discussions and V. Vinokur for his critical reading of the paper and
discussions. V.T. thanks DST, India, for a Swarnajayanti grant (DST/SJF/PSA-0212012-13)
and Argonne National Laboratory, USA, where a significant part of
this work was completed.

\begin{appendix}
 \subsection{RG equations and phase diagram of the disordered XY model}

In this section we show the essential steps followed for obtaining
the phase diagram of the two-dimensional XY model with phase disorder.
A comprehensive study can be found in Ref. \cite{carpentiernucphysb}.

The partition function of the replicated Coulomb gas with \textit{m-vector}
charges after averaging over the bare disorder is

\[
\overline{Z^{m}}=1+\sum_{p=2}^{\infty}\sum_{\mathbf{n}_{1},\ldots,\mathbf{n}_{p}}\int_{|\mathbf{r}_{i}-\mathbf{r}_{j}|>a_{0}}\exp(-\beta H^{(m)}[\mathbf{n},\mathbf{r}]),
\]
where the sum is over all distinct neutral configurations and

\[
\beta H^{(m)}=\sum_{i\not=j}K_{ab}n_{i}^{a}\ln\left(\frac{|\mathbf{r}_{i}-\mathbf{r}_{j}|}{a_{0}}\right)n_{j}^{b}+\sum_{i}\ln Y[\mathbf{n}_{i}].
\]
Here, $Y[\mathbf{n}]=\exp(-n^{a}\gamma K_{ab}n^{b}),$ where $K_{ab}=\beta J\delta_{ab}-\sigma\beta^{2}J^{2}.$
Significant contribution to the partition function only comes from
charges $\pm1,0$ and hence we restrict to these. We increase the
hard core cutoff $a_{0}\rightarrow a_{0}e^{(dl)}$ and retain the
original form of the partition function in terms of scale dependent
coupling constants $(K_{l})_{ab}$ and fugacities $Y_{l}[\mathbf{n}]$.
To $O(Y[\mathbf{n}]^{2})$, we obtain the following RG flow equations\cite{carpentiernucphysb}:

\begin{align}
\partial_{l}(K_{l}^{-1})_{ab} & =2\pi^{2}\sum_{\mathbf{n}\not=0}n^{a}n^{b}Y[\mathbf{n}]Y[-\mathbf{n}]\label{eq:renorm_coupling}\\
\partial_{l}Y[\mathbf{n}\ne0] & =(2-n^{a}K_{ab}n^{b})Y[\mathbf{n}]+\sum_{{\bf n}^{'}\ne0,{\bf n}}\pi Y[\mathbf{n}^{'}]Y[\mathbf{{\bf n}-n}^{'}]\label{eq:renorm_fugacity}
\end{align}

Equation(\ref{eq:renorm_coupling}) comes from the annihilation of
dipoles of opposite vector charges in the annulus $a_{0}<|r_{i}-r_{j}|<a_{0}e^{dl}$.
It gives the renormalization of the interaction and of the disorder.
Simple rescaling gives the first part of equation (\ref{eq:renorm_fugacity}).
The second part comes from the possibility of fusion of two replica
vector charges upon coarse graining. Some examples of fusion are given
below.

\begin{widetext} 
\[
\left(\begin{array}{c}
\vdots\\
+1\\
\vdots\\
+1\\
\vdots\\
0\\
\vdots
\end{array}\right)+\left(\begin{array}{c}
\vdots\\
0\\
\vdots\\
0\\
\vdots\\
-1\\
\vdots
\end{array}\right)\rightarrow\left(\begin{array}{c}
\vdots\\
+1\\
\vdots\\
+1\\
\vdots\\
-1\\
\vdots
\end{array}\right),\left(\begin{array}{c}
\vdots\\
+1\\
\vdots\\
0\\
\vdots\\
0\\
\vdots
\end{array}\right)+\left(\begin{array}{c}
\vdots\\
0\\
\vdots\\
+1\\
\vdots\\
0\\
\vdots
\end{array}\right)\rightarrow\left(\begin{array}{c}
\vdots\\
+1\\
\vdots\\
+1\\
\vdots\\
0\\
\vdots
\end{array}\right),\left(\begin{array}{c}
\vdots\\
+1\\
\vdots\\
0\\
\vdots\\
1\\
\vdots
\end{array}\right)+\left(\begin{array}{c}
\vdots\\
-1\\
\vdots\\
0\\
\vdots\\
0\\
\vdots
\end{array}\right)\rightarrow\left(\begin{array}{c}
\vdots\\
0\\
\vdots\\
0\\
\vdots\\
1\\
\vdots
\end{array}\right),
\]

\end{widetext}

Replica permutation symmetry, which we will assume here and which
is preserved by the RG, together with $n^{a}=0,\pm1$ implies that
$Y[\mathbf{n}]$ depends only on the numbers $n_{+}$ and $n_{-}$
of $+1/-1$ components of $\mathbf{n}$. We parameterize $Y[\mathbf{n}]$
by introducing a function of two arguments $\Phi(z_{+},z_{-})$, where
$z_{\pm}({\bf r})=\exp(\pm\beta v_{{\bf r}}),$ such that: 
\begin{equation}
Y[\mathbf{n}]=\left<z_{+}^{n_{+}}z_{-}^{n_{-}}\right>_{\Phi}
\end{equation}
where we denote $<A>_{\Phi}=\int dz_{+}dz_{-}A\Phi(z_{+},z_{-})$.
After some manipulations \cite{carpentiernucphysb}, in the limit $m\rightarrow0$,
we can write eq(\ref{eq:renorm_fugacity}) in terms of, $P=\phi/(\int_{z_{+},z_{-}>0}\phi)$,
which can be interpreted as a probability distribution, as

\begin{widetext}

\begin{equation}
\partial_{l}P(z_{+},z_{-})=\mathcal{O}P-2P(z_{+},z_{-})+2\left<\delta\left(z_{+}-\frac{z_{+}^{'}+z_{+}^{''}}{1+z_{-}^{'}z_{+}^{''}+z_{+}^{'}z_{-}^{''}}\right)\delta\left(z_{-}-\frac{z_{-}^{'}+z_{-}^{''}}{1+z_{-}^{'}z_{+}^{''}+z_{+}^{'}z_{-}^{''}}\right)\right>_{P^{'}P^{''}},\label{eq:p_renorm}
\end{equation}

\end{widetext}

where, $\mathcal{\mathcal{O}}=\beta J(2+z_{+}\partial_{z_{+}}+z_{-}\partial_{z_{-}})+\sigma(\beta J)^{2}(z_{+}\partial_{z_{+}}-z_{-}\partial_{z_{-}})^{2}.$
The $m\rightarrow0$ limit of eq(\ref{eq:renorm_coupling}) similarly
yields,

\begin{align}
T\frac{dJ^{-1}}{dl} & =8\left<\frac{z_{+}^{'}z_{-}^{''}+z_{-}^{'}z_{+}^{''}+4z_{+}^{'}z_{-}^{''}z_{-}^{'}z_{+}^{''}}{(1+z_{+}^{'}z_{-}^{''}+z_{-}^{'}z_{+}^{''})^{2}}\right>_{PP}\label{eq:J_renorm}\\
\frac{d\sigma}{dl} & =8\left<\frac{(z_{+}^{'}z_{-}^{''}-z_{-}^{'}z_{+}^{''})^{2}}{(1+z_{+}^{'}z_{-}^{''}+z_{-}^{'}z_{+}^{''})^{2}}\right>_{PP}\label{eq:sigma_remorm}
\end{align}

Equations (\ref{eq:p_renorm}),(\ref{eq:J_renorm}) and (\ref{eq:sigma_remorm})
form the complete set of RG equations.

Numerical study\cite{carpentiernucphysb} of the RG equations indicate
the existence of an XY phase at low temperatures and below some critical
disorder. Guided by the RG flow observed numerically within and near
the boundaries of the XY phase, we can approximate the full RG equations
by a simpler equation involving only the single fugacity distribution,
$P_{l}(z)=\int dz_{+}P_{l}(z_{+},z)=\int dz_{-}P_{l}(z,z_{-})$. In
the low T regime, the distribution $P_{l}(z_{+},z_{-})$ is broad
and the physics is dominated by rare favorable regions $(z_{+}\sim1$
or $z_{-}\sim1)$. Here we identify a parameter that allows to organise
perturbation theory as: $P_{l}(1)\equiv P_{l}(z\sim1)\sim P_{l}(z_{+}\sim1,z_{-}\sim0)=P_{l}(z_{+}\sim0,z_{-}\sim1)$We
also observe that $P_{l}(1,1)\equiv P_{l}(z_{+}\sim1,z_{-}\sim1)\sim P_{l}(1)^{2}$.
Using these we can see schematically the RG equation (\ref{eq:p_renorm})
as a correction to $P_{l}(1)$ of order $P_{l}(1)$ by the first term
and order $P_{l}(1)^{2}$ by the second term; in RG equation (\ref{eq:J_renorm}),(\ref{eq:sigma_remorm})
as a correction to order $P_{l}(1)^{2}$ to $J_{l}$ and $\sigma_{l}$.
Again working to order $P_{l}(1)^{2}$, we see that the denominators
in the delta functions in (\ref{eq:p_renorm}) could be neglected.
This approximation also simplifies equations (\ref{eq:J_renorm})
and (\ref{eq:sigma_remorm}).

Introducing 
\begin{equation}
G_{l}(x)=1-\int_{-\infty}^{\infty}du\tilde{P}_{l}(u)exp(-e^{\beta(u-x+E_{l})}),\label{eq:param_g}
\end{equation}
where $u=1/\beta\ln(z)$ and $E_{l}=\int_{0}^{l}J(l^{'})dl^{'}$,
we see that (\ref{eq:p_renorm}) can be written as $\frac{1}{2}\partial_{l}G=\frac{\sigma J^{2}}{2}\partial_{x}^{2}G+G(1-G)$.
If $\sigma$ and $J$ are $l$ independent we identify the above with
Kolmogorov-Petrovskii-Piscounov (KPP) equation, whose general form
is , $\frac{1}{2}\partial_{l}G=D\partial_{x}^{2}G+f(G)$, where $D$
is a constant and $f$ satisfies $f(0)=f(1)=0$, $f$ positive between
0 and 1 and $f^{'}(0)=1,f^{'}(G)\leq1$ between 0 and 1. Since at
large $l$, both $J$ and $\sigma$ converge and effectively becomes
$l$ independent, we see that we can use results from the study of
KPP equation in our case at large $l$.

For a large class of initial conditions, the solutions of the KPP
equation are known to converge uniformly towards traveling wave solutions
of the form: $G_{l}(x)\rightarrow h(x-m_{l}).$ The velocity of the
wave is given by $c=\lim_{l\rightarrow\infty}\partial_{l}m_{l}$.
A theorem due to Bramson\cite{bramson} shows that the asymptotic
traveling wave is determined by the behavior at $x\rightarrow\infty$
of the initial condition $G_{l=0}(x)$ in the following manner. If
$G_{l=0}(x)$ decays faster than $e^{-\mu x}$ where $\mu=1/\sqrt{D}$,
then $c=\sqrt{D}$. If $G_{l=0}(x)$ decays slower than $e^{-\mu x}$
where $\mu<1/\sqrt{D}$, then $c=2(D\mu+\mu^{-1})$. The parameterization(\ref{eq:param_g})
implies that the distribution $\tilde{P}_{l}(u)$ itself converges
to a traveling front solution 
\begin{equation}
\tilde{P}_{l}(u)\rightarrow_{l\rightarrow\infty}\tilde{p}(u-X_{l})\mbox{ , }X_{l}=m_{l}-E_{l}.
\end{equation}
Since $\partial_{l}E_{l}\rightarrow_{l\rightarrow\infty}J_{R}$, we
see that the asymptotic velocity of the front of $\tilde{P}_{l}(u)$
is $c-J_{R}$, where $c$ is the KPP front velocity. The center of
the front corresponds to the maximum of the distribution $\tilde{P}(u)$.

The asymptotic velocity clearly decides the phase of the system: since
we start with a distribution peaked at some small $z$, if the velocity
is positive, then $P_{l}(1)$ will increase and this would imply that
the system is in the disordered phase. On the other hand negative
velocity implies that the system is in the XY phase. The velocity
vanishes at the phase boundary. By construction, the initial condition
$G_{l=0}(x)$ decays for large x as $<z>_{P_{0}}e^{-\beta x}$. Hence
we identify $\mu=\beta$. Based on the results discussed above about
the front velocity selection in KPP equation we can conclude the following
about the phase diagram of the model:

(a) For $T>T_{g}=J_{R}\sqrt{\sigma_{R}/2}$, $c=T\left(2+\frac{\sigma_{R}J_{R}^{2}}{T^{2}}\right)$.
Thus here the XY phase would exist for 
\begin{equation}
2-\frac{J_{R}}{T}+\frac{\sigma_{R}J_{R}^{2}}{T^{2}}<0.
\end{equation}

(b) For $T\leq T_{g}$, $c=J_{R}\sqrt{8\sigma_{R}}$. Thus here the
XY phase would exist for $\sigma_{R}<\sigma_{c}=\frac{1}{8}$.

\textit{Critical behavior at zero temperature: }The zero temperature
phase transition from the XY phase to the disordered phase occurs
at $\sigma_{R}=1/8$. The center of the front is located at $u=X_{l}$
near the transition. It follows from \cite{bramson} that, $X_{l}\approx(4\sqrt{D}-J)l-3/2\sqrt{D}\ln l+X_{0}.$Hence
in the critical region to leading order, we get,

\begin{equation}
\partial_{l}X_{l}\sim4\sqrt{D}-J-\frac{3\sqrt{D}}{2l}.\label{eq:renorm_X}
\end{equation}

After some manipulations the RG equations for $J$ and $\sigma$ in
the critical region reads,

\begin{align*}
\partial_{l}(J^{-1}) & =k\int du\tilde{p_{l}}(u-X_{l})\tilde{p_{l}}(-u-X_{l})\\
\partial_{l}\sigma & =k\int_{u+u^{'}>-2X_{l}}\tilde{p_{l}}(u)\tilde{p_{l}}(u^{'}),
\end{align*}

where $k$ is some constant. Using the asymptotic form of $\tilde{p_{l}}(u)$discussed
in \cite{bramson} and working upto leading order in $(\sigma-\sigma_{c})$,
we can simplify the above equations to get,

\begin{align*}
\partial_{l}(J^{-1}) & \sim\frac{C}{\sqrt{D}}X_{l}^{3}\exp\left(\frac{2X_{l}}{\sqrt{D}}\right)\\
\partial_{l}\sigma & \sim CX_{l}^{3}\exp\left(\frac{2X_{l}}{\sqrt{D}}\right),
\end{align*}

where $C$ is a constant. To estimate the form of correlation length,
we first introduce the small parameter,$g_{l}=\exp(X_{l}/\sqrt{D}).$
Then (\ref{eq:renorm_X}) reads,

\[
\partial_{l}g\sim\left(16(\sigma-\sigma_{c})-\frac{3}{2l}\right)g
\]

Now starting away from criticality, $\epsilon=\sigma_{c}-\sigma_{R}>0$,
we find, $g_{l}\sim l^{-3/2}\exp(16\epsilon l).$ Identifying the
correlation length $\xi$ as when $g_{\xi}\sim1$, we find,

\[
\xi\sim\exp\left(\frac{b}{|\sigma-\sigma_{c}|}\right),
\]

where $b$ is some constant. We then see that the universality class
of this transition is clearly different from the KT universality class.

\begin{figure}
\includegraphics[width=7.5cm]{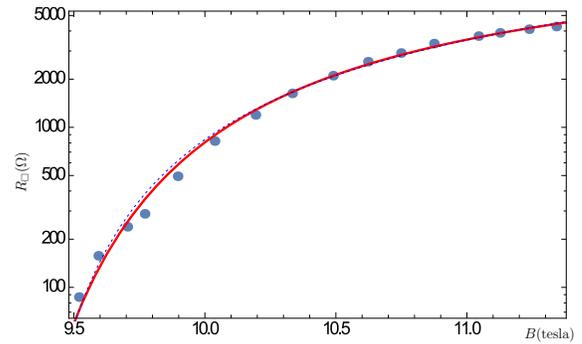}

\caption{\label{fig:ktab}Sheet resistance $R_{\Box}$ of homogeneously-disordered
InO$_{x}$ thin films as a function of perpendicular magnetic field
$B$ near the field driven SIT (data extracted from Ref. \cite{kapitulnik}).
The fits are to a Kosterlitz-Thouless (KT) law $R_{\Box}=5.02\times10^{4}e^{-16/\sqrt{B^{2}-9.2^{2}}}$
(solid red curve) and the non-KT law $R_{\Box}=1.53\times10^{4}e^{-61.2/(B^{2}-8.9^{2})}$
(blue dashed curve). The KT transition is driven by thermal phase
fluctuations while the non-KT transition is driven by phase frustration.
Both the laws fit the data equally well; however the pre-factor of
the exponential, which represents the high-field sheet resistance,
is a more reasonable number in the non-KT case since in the actual
data, the peak value of resistance is of a comparable order.}
\end{figure}

\subsection{Comparison of Kosterlitz-Thouless (KT) and non-KT scaling with experiments}

In Fig.\ref{fig:ktab}, we show the sheet resistance $R_{\Box}$ vs.
magnetic field data near a field-driven SIT in a homogeneously-disordered
InO$_{x}$ thin film from Ref. \cite{kapitulnik}, and attempt fits
of this data to the Kosterlitz-Thouless (KT) behavior ($R_{\Box}=R_{0}e^{-1/\sqrt{B-B_{KT}}}$)
and the non-KT behavior ($R_{\Box}=R_{0}e^{-1/(B-B_{AB})}$) obtained
in this Letter. It is difficult to say which of these two laws describes
the data better; however, we argue that the non-KT fit might be a
bit better on account of a more reasonable value for the high-field
resistance $R_{0}.$
\end{appendix}

\end{acknowledgments}
\bibliographystyle{apsrev4-1}
\bibliography{ref}

\end{document}